\documentclass[11pt]{article}

\usepackage[utf8]{inputenc}
\usepackage[T1]{fontenc}
\usepackage{amsmath,amssymb}
\usepackage{textcomp}
\usepackage[margin=2.5cm]{geometry}
\usepackage{graphicx}
\usepackage[colorlinks=true, allcolors=blue]{hyperref}
\usepackage{booktabs}
\usepackage{array}
\usepackage{longtable}
\usepackage{tikz}
\usepackage{xcolor}
\usepackage{colortbl}
\usepackage{enumitem}
\usetikzlibrary{arrows.meta, positioning, shapes.geometric}

\title{When Not to Automate: A Formal Protocol for Human Preservation in AI-Optimized Organizations}
\author{
  Jairo Rodr\'iguez Arias$^1$ \quad Spyridon Chouliaras$^1$ \quad Jos\'e Manuel de la Chica$^1$ \\
  \small $^1$Santander AI Lab
}
\date{July 2026}

\begin{document}
\maketitle

\begin{abstract}
Standard automation ROI misses four categories of systemic risk---tacit knowledge erosion, resilience reduction, regulatory exposure, and socio-institutional capital degradation---that affect long-term organizational performance. PHP-AIO (Protocol for Human Preservation in AI-Optimized Organizations) is a five-gate sequential decision protocol with a final composite check that quantifies these unpriced systemic risks at the role level and produces auditable automation decisions. A closed-form automation-debt measure ($\rho(P)$) formalises how role-level decisions accumulate across multi-step processes; its warning is neutralised only by a regulator-mandated human-in-the-loop anchor. Applied to stylised profiles of representative internal roles, PHP-AIO produces distinct outcomes---automate, augment, hybrid, and preserve---for candidates that standard cost-benefit analysis would uniformly automate. Threshold sensitivity analysis confirms the gate decisions are robust to upward perturbations of at least 14\% in three of four representative cases.

\end{abstract}

\section{The Problem: Automation's Blind Spot}

PHP-AIO is a five-gate decision protocol that quantifies four categories of unpriced systemic risk in automation decisions and produces auditable outcomes---automate, preserve, augment, or hybrid---each with prescribed governance actions.

The protocol addresses a structural blind spot in how organizations decide to automate. Standard ROI models optimize for cost per transaction, processing time, and error rates \cite{brynjolfsson1993productivity, acemoglu2019automation}. They miss four long-term costs: the irreversible loss of tacit knowledge accumulated over decades, the reduction of organizational redundancy that buffers AI failures, the rising regulatory exposure as jurisdictions mandate human oversight, and the erosion of the public trust capital that financial institutions depend on \cite{brynjolfsson2022turing}.

We frame these four costs under a single organising principle: \emph{automation sovereignty}---an organisation's capacity to retain strategic autonomy over its consequential decisions as AI capabilities advance. We introduce the term automation sovereignty in this paper as an organising frame for the four risk dimensions; it does not appear under this label in prior work, but it consolidates concerns already present in Mittelstadt's principles-to-practice gap \cite{mittelstadt2019principles} and Wagner's analysis of accountability shifts under automated decision systems \cite{wagner2019liable}. The automation-debt construct of Section~\ref{sec:automation-debt} is its process-level operationalisation.

Tacit-knowledge erosion (TKE) compromises sovereignty by removing the institutional memory required to recover from automation failure; resilience reduction (RR), by creating correlated dependencies on the systems that automate decisions; regulatory exposure (RE), by transferring decision-making authority to actors outside the firm when oversight is mandated retroactively; and socio-institutional capital degradation (SCD), by depleting the trust that makes consequential decisions enforceable. PHP-AIO operationalises automation sovereignty as a continuous risk-and-debt accounting discipline, not a binary automate-or-not choice; automation debt (introduced in Section~\ref{sec:automation-debt}) is its process-level density signal across multi-step processes.

We call these four categories systemic risks. Scholars have documented tacit-knowledge erosion \cite{bainbridge1983ironies, collins2010tacit}, automation-induced complacency and overreliance on algorithmic recommendations \cite{parasuraman2010complacency, bucinca2021trust, cummings2004automation}, organizational brittleness \cite{perrow1984normal, taleb2012antifragile}, and the principles-to-practice gap in AI ethics \cite{floridi2020design, mittelstadt2019principles}. Recent surveys consolidate the human-in-the-loop design space \cite{mosqueirarey2023human}. The gap is not awareness, it is operational: no formal protocol incorporates these risks into the automation decision itself. Existing governance frameworks---the US NIST AI Risk Management Framework \cite{nist2023airmf}, the EU AI Act \cite{euaiact2024, veale2021demystifying}, and human-in-the-loop (HITL) oversight taxonomies---engage after the deployment-commitment decision. None decide whether to automate. Sociotechnical scholarship moreover warns that procedural human oversight and technical compliance alone do not guarantee meaningful control when system design embeds the wrong abstractions \cite{selbst2019fairness, wagner2019liable, green2022flaws, shneiderman2022human}.

\paragraph{Where PHP-AIO sits.} Table~\ref{tab:comparison} contrasts PHP-AIO against the four closest governance frameworks. The NIST AI Risk Management Framework \cite{nist2023airmf} is voluntary cross-sector guidance organised around four functions (govern, map, measure, manage); it does not prescribe thresholds or decisions. The EU AI Act conformity assessment \cite{euaiact2024, veale2021demystifying} mandates pre-market and post-market checks for high-risk systems---human-oversight requirements, Annex IV technical documentation, and post-market monitoring---but only once a system is classified as high-risk. FAccT-style algorithmic impact assessments \cite{metcalf2021algorithmic, moss2021assembling} produce structured analyses of fairness, discrimination, and sociotechnical context, ranging from qualitative narratives to participatory deliberation processes; outputs are typically used to inform deployment governance rather than as binary decision filters. AI Risk Profiles \cite{sherman2024risk}, the closest pre-decision competitor, provide pre-deployment risk-profile disclosure organised around a taxonomy of AI risks but use ordinal severity levels rather than formal scoring and do not produce a deployment decision. Three axes distinguish PHP-AIO from each comparator: (i) formal scoring---PHP-AIO produces explicit numerical gate scores rather than narrative or ordinal outputs; (ii) an explicit decision output among four outcomes (automate / augment / hybrid / preserve); and (iii) jurisdictional granularity through country-specific multipliers in the RE formula. Where PHP-AIO also differs in timing, it acts pre-decision---before the deployment-commitment choice is taken---whereas the other frameworks engage once a deployment candidate exists \cite{raisch2021artificial}.

\begin{table}[htbp]
\centering
\caption{PHP-AIO relative to the closest governance frameworks. Pin-cite mapping: NIST Timing \& Multi-dim. $\rightarrow$ NIST AI RMF 1.0 \S3.1 lifecycle framing and \S3.2 trustworthy characteristics; EU AI Act Timing $\rightarrow$ Reg. 2024/1689 Art. 43 (pre-market conformity assessment) + Art. 72 (post-market monitoring); EU AI Act Multi-dim. $\rightarrow$ Annex III high-risk classes; AI Risk Profiles $\rightarrow$ \cite{sherman2024risk} 4-part taxonomy. The four comparators address different points in the AI lifecycle; PHP-AIO acts upstream of the deployment-commitment decision itself.}
\label{tab:comparison}
\resizebox{\textwidth}{!}{%
\begin{tabular}{@{}l p{2.6cm} p{2.8cm} p{2.4cm} p{2.8cm} p{2.6cm}@{}}
\toprule
 & \textbf{NIST AI RMF} \cite{nist2023airmf} & \textbf{EU AI Act} \cite{euaiact2024,veale2021demystifying} & \textbf{FAccT IA} \cite{metcalf2021algorithmic,moss2021assembling} & \textbf{AI Risk Profiles} \cite{sherman2024risk} & \textbf{PHP-AIO} (this paper) \\
\midrule
Timing & lifecycle (all phases) & pre-market + post-market & pre-deployment & pre-deployment & pre-decision \\
Output type & recommendations & binary conformity & narrative impact & risk-profile disclosure & score + decision \\
Formal scoring & $\times$ & $\times$ & $\times$ & ordinal (taxonomy) & \checkmark~(5 gates) \\
Multi-dim. risk & 7 trust characteristics & Annex III high-risk areas & varies & taxonomy of AI risks & 4 dim.\ + composite \\
Jurisdiction-aware & $\times$ & implicit (EU) & $\times$ & $\times$ & \checkmark~(10 countries) \\
\bottomrule
\end{tabular}%
}
\end{table}

\section{Four Dimensions of Unpriced Risk}
\label{sec:dimensions}

Each of the four risk dimensions captures a distinct mechanism through which automation can damage organizational performance: TKE captures the irreversible loss of judgment that cannot be re-encoded once a role is removed; RR captures the correlated fragility introduced when automated systems share failure modes; RE captures the option-value cost of reversing automation if regulation later mandates human oversight; and SCD captures the depletion of the trust capital that makes consequential decisions enforceable.

\paragraph{Common scale convention.} Every sub-dimension is first computed from its primary inputs (Likert scales, boolean dichotomies, monetary thresholds) and then affinely rescaled to $[0,1]$ using its effective input range. If a sub-dimension $s$ takes raw values in $[s_{\min}, s_{\max}]$, its normalised score is $(s - s_{\min})/(s_{\max} - s_{\min})$, clipped to $[0,1]$. The rescaling guarantees that a value of 0.5 means halfway through the risk range consistently across all twelve sub-dimensions, and that the four composite scores (TKE, RR, RE, SCD) are directly comparable on the same axis. The bullets in each subsection report the raw construction for transparency.

\paragraph{On weights.} All four risk dimensions take the same form---a weighted sum of three normalised sub-dimensions, with the weights summing to one. The per-dimension weights reported alongside each formula below (0.4/0.3/0.3 for TKE; 0.35/0.35/0.30 for RR and SCD; 0.4/0.4/0.2 for RE) are an expert configuration, not a derived constant. They encode a governance choice about the relative importance of each sub-dimension and are expected to vary by industry, use case, and risk appetite (in healthcare, for instance, RE typically over-weights its CQ sub-component). The values reported here are the financial-services baseline, elicited through internal Santander AI Lab consensus over Q1 2026 and informed by \cite{cummings2004automation} on automation bias and \cite{mittelstadt2019principles} on algorithmic-governance practice. The four gates use a uniform threshold of 0.70, so the same risk score is treated identically across dimensions.

\subsection{Tacit Knowledge Erosion (TKE)}

Tacit knowledge is the embodied, experience-derived judgment practitioners develop through years of domain-specific practice \cite{nonaka1995knowledge, collins2010tacit}; explicit documentation cannot fully capture it \cite{rintakahila2018consequences, endsley2017here}. TKE combines three sub-dimensions---codifiability ($\kappa$), irreversibility (IR), and criticality (CM)---as a weighted sum:

\begin{equation}
\mathrm{TKE}(r) = 0.4 \cdot (1 - \kappa) + 0.3 \cdot \mathrm{IR} + 0.3 \cdot \mathrm{CM}
\label{eq:tke}
\end{equation}

The $(1-\kappa)$ term flips codifiability: a fully codifiable role contributes 0 to that term. Example: a senior role with 5 years of experience, high expert judgment, and high error impact yields $\mathrm{TKE} \approx 0.56$ (Appendix~\ref{app:worked-examples}; sub-dimension construction in Appendix~\ref{app:subdimensions}).

\subsection{Resilience Reduction (RR)}

AI systems fail in correlated ways under novel, adversarial, or high-ambiguity conditions \cite{mcgregor2021preventing}; human roles provide a non-correlated resilience layer whose removal makes organisations brittle in exactly those regimes \cite{hollnagel2011four, woods2018theory}. Pairing humans with AI does not automatically restore resilience: gains depend on careful role and explanation design \cite{bansal2021does}. RR combines three sub-dimensions---Recovery (RC; manual fallback and downtime tolerance), Adversarial (AD; novel-threat detection and human-escalation rate), and Correlation (CR; count of dependent automated systems)---as a weighted sum:

\begin{equation}
\mathrm{RR}(r) = 0.35 \cdot \mathrm{RC} + 0.35 \cdot \mathrm{AD} + 0.30 \cdot \mathrm{CR}
\label{eq:rr}
\end{equation}

Higher values mean greater resilience risk. Example: a loan-approval role with no manual fallback, 24h downtime tolerance, low novel-threat detection, and three dependent downstream systems yields $\mathrm{RR} \approx 0.61$ (full derivation in Appendix~\ref{app:worked-examples}, construction in Appendix~\ref{app:subdimensions}).

\subsection{Regulatory Exposure (RE)}

Regulatory frameworks across jurisdictions are converging on mandatory human oversight for consequential automated decisions \cite{smuha2021race, euaiact2024}. Automation decisions made today carry regulatory option value---the cost of reversing automation if future regulation requires human review. RE combines three sub-dimensions---Consequentiality (CQ; customer impact, credit-decision involvement, sensitive-data handling), Jurisdictional (JU; audit-body count, HITL mandates, and precedent sanctions, multiplied by a country factor in $\{0.8, 0.9, 1.0, 1.1, 1.2\}$ across the ten countries with explicit multipliers), and Reversal (RV; cost to reintroduce human oversight, capped at \texteuro500K)---as a weighted sum:

\begin{equation}
\mathrm{RE}(r) = 0.4 \cdot \mathrm{CQ} + 0.4 \cdot \mathrm{JU} + 0.2 \cdot \mathrm{RV}
\label{eq:re}
\end{equation}

Example: a UK loan-approval role (sensitive data, HITL mandatory, three audit bodies, \texteuro200K reversal cost) yields $\mathrm{RE} \approx 0.85 \rightarrow$ hybrid (full derivation in Appendix~\ref{app:worked-examples}, construction in Appendix~\ref{app:subdimensions}).

\subsection{Socio-Institutional Capital Degradation (SCD)}

Financial institutions operate on trust \cite{zucker1986production, north1990institutions}. Certain human roles serve as trust anchors: their presence signals care, accountability, and human agency \cite{putnam2000bowling}. Replacing them with automation triggers trust deficits that resist advance measurement but turn catastrophic when they materialize. SCD combines three sub-dimensions---Relational (RL; customer-interaction depth and explicit human expectation \cite{dietvorst2015algorithm, logg2019algorithm}), Legitimacy (LG; brand-perception impact and public visibility), and Labor (LB; team-morale impact and headcount in the role)---as a weighted sum:

\begin{equation}
\mathrm{SCD}(r) = 0.35 \cdot \mathrm{RL} + 0.35 \cdot \mathrm{LG} + 0.30 \cdot \mathrm{LB}
\label{eq:scd}
\end{equation}

Example: a retail in-branch advisor (in-person interaction, customers expect a human, 30 employees, brand impact 5/5) yields $\mathrm{SCD} \approx 0.90 \rightarrow$ augment (full derivation in Appendix~\ref{app:worked-examples}, construction in Appendix~\ref{app:subdimensions}).

\section{The Five-Gate Decision Protocol}
\label{sec:protocol}

\begin{figure}[htbp]
\centering
\begin{tikzpicture}[
  node distance=1.1cm and 1.6cm,
  gate/.style={rectangle, draw, rounded corners, minimum width=2.6cm, minimum height=1cm, align=center, fill=blue!8},
  outcome/.style={rectangle, draw, rounded corners, minimum width=2.2cm, minimum height=0.9cm, align=center, fill=orange!12},
  auto/.style={rectangle, draw, rounded corners, minimum width=2.2cm, minimum height=0.9cm, align=center, fill=green!15},
  >=Stealth
]
\node[gate] (g1) {G1 Net Benefit\\$\geq 15\%$};
\node[gate, right=of g1] (g2) {G2 TKE\\$< 0.70$};
\node[gate, right=of g2] (g3) {G3 RR\\$< 0.70$};
\node[gate, right=of g3] (g4) {G4 RE\\$< 0.70$};
\node[gate, right=of g4] (g5) {G5 SCD\\$< 0.70$};
\node[gate, right=of g5] (comp) {Composite\\$< 0.60$};
\node[auto, right=of comp] (automate) {automate};

\node[outcome, below=of g1] (p1) {preserve};
\node[outcome, below=of g2] (p2) {preserve};
\node[outcome, below=of g3] (p3) {preserve};
\node[outcome, below=of g4] (h4) {hybrid};
\node[outcome, below=of g5] (a5) {augment};
\node[outcome, below=of comp] (hcomp) {hybrid};

\draw[->] (g1) -- node[above, font=\scriptsize]{pass} (g2);
\draw[->] (g2) -- node[above, font=\scriptsize]{pass} (g3);
\draw[->] (g3) -- node[above, font=\scriptsize]{pass} (g4);
\draw[->] (g4) -- node[above, font=\scriptsize]{pass} (g5);
\draw[->] (g5) -- node[above, font=\scriptsize]{pass} (comp);
\draw[->] (comp) -- node[above, font=\scriptsize]{pass} (automate);

\draw[->, dashed] (g1) -- node[right, font=\scriptsize]{fail} (p1);
\draw[->, dashed] (g2) -- node[right, font=\scriptsize]{fail} (p2);
\draw[->, dashed] (g3) -- node[right, font=\scriptsize]{fail} (p3);
\draw[->, dashed] (g4) -- node[right, font=\scriptsize]{fail} (h4);
\draw[->, dashed] (g5) -- node[right, font=\scriptsize]{fail} (a5);
\draw[->, dashed] (comp) -- node[right, font=\scriptsize]{fail} (hcomp);
\end{tikzpicture}
\caption{Five-gate protocol with its composite check. Solid arrows denote a passed gate; dashed arrows denote the failure outcome routed by each gate.}
\label{fig:gate-flow}
\end{figure}
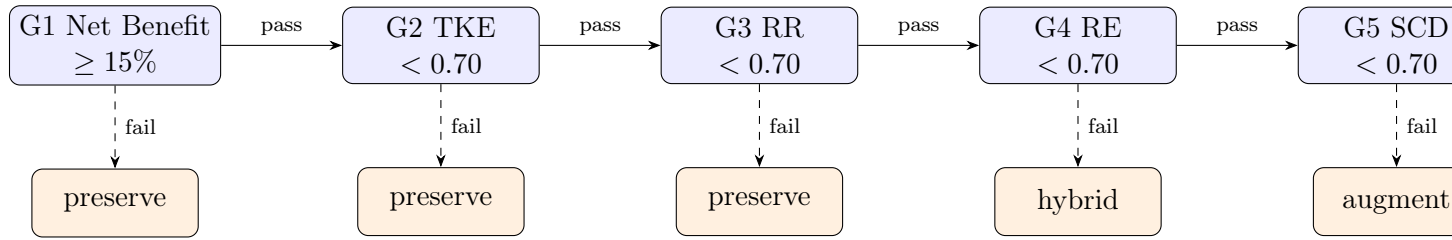

PHP-AIO evaluates automation candidates through five sequential gates. A role must clear each gate before reaching the next. Gate 1 is a net-benefit pre-filter: automation must deliver a minimum positive net benefit to justify risk assessment. The remaining four gates correspond to the risk dimensions defined in Section~\ref{sec:dimensions}. The sequential structure assesses irreversible risks first. Table~\ref{tab:gates} specifies the gate questions, thresholds, and failure outcomes; Figure~\ref{fig:gate-flow} renders the same protocol as a decision flow. Each gate's failure routes the role to a specific outcome: G1, G2, or G3 $\rightarrow$ preserve (keep the human role, protecting against irreversible loss); G4 $\rightarrow$ hybrid (AI executes, human supervises, remediating regulatory exposure); G5 $\rightarrow$ augment (AI assists, human decides, protecting trust capital). A role that clears all five individual gates is then subjected to a final composite check: a weighted average of the four risk dimensions that must remain below a tighter threshold ($\geq 0.60$ routes to hybrid; the per-gate threshold is 0.70). The composite catches roles whose four dimensions individually sit just below their thresholds but whose accumulated score signals systemic moderate risk---a failure mode that single-gate checks miss by construction. Roles that clear all five gates and the composite check yield automate; the formal definition, weights, and rationale for the asymmetric threshold are in Section~\ref{sec:composite}.

\paragraph{Why a cost threshold first.} Gate 1 is a cheap necessary filter, not a sufficiency claim. It reserves the costly risk analysis of Gates 2--5 for candidates with at least a baseline economic case. Cost savings does not decide the outcome---the four qualitative gates do.

\begin{table}[htbp]
\centering
\caption{The five-gate protocol with its composite check.}
\label{tab:gates}
\begin{tabular}{@{}c l p{5cm} l l@{}}
\toprule
\textbf{Gate} & \textbf{Dimension} & \textbf{Question} & \textbf{Threshold} & \textbf{If Failed} \\
\midrule
G1 & Net Benefit & Does automation produce $\geq 15\%$ net benefit? & Net benefit $\geq 15\%$ & preserve \\
G2 & TKE & Is tacit knowledge risk below threshold? & $\mathrm{TKE} < 0.70$ & preserve \\
G3 & RR & Is resilience impact acceptable? & $\mathrm{RR} < 0.70$ & preserve \\
G4 & RE & Is regulatory exposure manageable? & $\mathrm{RE} < 0.70$ & hybrid \\
G5 & SCD & Is social capital impact acceptable? & $\mathrm{SCD} < 0.70$ & augment \\
Composite & All four dims & Is accumulated moderate risk acceptable? & $\mathrm{PHP\text{-}AIO} < 0.60$ & hybrid \\
\bottomrule
\end{tabular}
\end{table}

Gates 2--3 protect against irreversible damage (failure $\rightarrow$ preserve); Gate 4 addresses regulatory risk through oversight (failure $\rightarrow$ hybrid); Gate 5 protects the customer relationship (failure $\rightarrow$ augment). Gate 1 measures direct cost savings by default; other monetised components (quality uplift, control strengthening, resilience, compliance, customer/employee experience) are admissible if each is named, sized and auditable. Threshold calibration, regulatory grounding, sensitivity analysis, and configurability per industry are discussed in Appendix~\ref{app:calibration}.

\subsection{Composite Score}
\label{sec:composite}

If a role passes all five gates, a composite score detects accumulated moderate risks:

\begin{equation}
\mathrm{PHP\text{-}AIO}(r) = w_1 \cdot \mathrm{TKE} + w_2 \cdot \mathrm{RR} + w_3 \cdot \mathrm{RE} + w_4 \cdot \mathrm{SCD}
\label{eq:composite}
\end{equation}

The financial-services configuration is $w_1 = 0.30$ (TKE), $w_2 = 0.25$ (RR), $w_3 = 0.30$ (RE), $w_4 = 0.15$ (SCD); weights sum to one and are configurable per domain. These values are an AI Lab Santander research configuration, elicited through internal expert consensus over Q1 2026; they do not represent an official Santander Group automation policy. The pairing $w_1 = w_3 = 0.30$ encodes the design principle that regulatory exposure and tacit-knowledge erosion are functionally equivalent in severity (both effectively irreversible). RR receives the middle weight $w_2 = 0.25$ because organisational redundancy and manual fallbacks are recoverable at high operational cost, and SCD the lowest $w_4 = 0.15$ because trust capital, although slow to rebuild, is the most recoverable of the four on a 3--5 year horizon and is partially absorbed by RE's CQ sub-component. The ordering is an editorial choice; neither EU regulatory frameworks nor banking-supervisory guidance currently rank automation-related risk categories at this granularity, leaving the weight assignment to institutional governance. If the composite $\geq 0.60$, the role receives a hybrid recommendation. The composite threshold is deliberately tighter than the per-gate 0.70: a role whose four dimensions each score, say, 0.65 clears every individual gate but yields a weighted average of $\approx 0.65 \geq 0.60$, signalling the death-by-a-thousand-cuts failure mode in which no single risk is acute yet the joint profile is unacceptable. The 0.60 value is calibrated against the four worked examples in Section~\ref{sec:examples}: it routes roles with one borderline-passing dimension (composite $\approx 0.45$--$0.55$) to automate, while flagging profiles with two or more borderline dimensions for human supervision. Single-gate decisions are robust to weight perturbations; composite-driven decisions are weight-sensitive by construction (Appendix~\ref{app:calibration} bounds the one composite-driven example of Section~\ref{sec:examples}).

\subsection{Cumulative Risk: Automation Debt}
\label{sec:automation-debt}

Task-level decisions stack: a process composed of subtasks individually safe to automate may still erode end-to-end resilience as dependencies multiply, manual fallbacks atrophy, and reversibility costs compound---a pattern that mirrors the accumulation dynamics of technical debt in software \cite{cunningham1992wycash} and in machine learning systems \cite{sculley2015hidden}. For a process $P$ with leaf subtasks $t_1, \ldots, t_n$, decisions $d_i$, and time weights $w_i = \mathrm{avg\_time\_minutes}(t_i)$, we define the time-weighted automation density

\begin{equation}
\rho(P) = \frac{\sum_{i=1}^{n} w_i \cdot \mathbb{1}[d_i = \mathrm{automate}]}{\sum_{i=1}^{n} w_i}
\label{eq:rho}
\end{equation}

\begin{figure}[htbp]
\centering
\begin{tikzpicture}[
  node distance=0.5cm and 0.5cm,
  leaf/.style={rectangle, draw, minimum width=1.7cm, minimum height=0.9cm, align=center, font=\scriptsize},
  root/.style={rectangle, draw, thick, minimum width=2.2cm, minimum height=0.9cm, align=center, font=\scriptsize, fill=gray!10},
  >=Stealth
]

\node[root] (rootA) {Loan\\Origination};
\node[leaf, below=1.3cm of rootA, xshift=-4.8cm, fill=green!20] (a1) {Doc digit.\\automate};
\node[leaf, right=0.3cm of a1, fill=green!20] (a2) {ID verify\\automate};
\node[leaf, right=0.3cm of a2, fill=green!20] (a3) {Credit pull\\automate};
\node[leaf, right=0.3cm of a3, fill=green!20] (a4) {Underwr.\\automate};
\node[leaf, right=0.3cm of a4, fill=green!20] (a5) {Offer gen.\\automate};
\draw (rootA) -- (a1); \draw (rootA) -- (a2); \draw (rootA) -- (a3); \draw (rootA) -- (a4); \draw (rootA) -- (a5);
\node[below=0.15cm of a3, font=\scriptsize, align=center] {\textbf{Scenario A}: $\rho(P)=5/5=1.00$\\no HITL anchor $\Rightarrow$ \textbf{debt warning}};

\node[root, right=4.5cm of rootA] (rootB) {Loan\\Origination};
\node[leaf, below=1.3cm of rootB, xshift=-4.8cm, fill=green!20] (b1) {Doc digit.\\automate};
\node[leaf, right=0.3cm of b1, fill=green!20] (b2) {ID verify\\automate};
\node[leaf, right=0.3cm of b2, fill=green!20] (b3) {Credit pull\\automate};
\node[leaf, right=0.3cm of b3, fill=yellow!30] (b4) {Underwr.\\hybrid\\(HITL anchor)};
\node[leaf, right=0.3cm of b4, fill=green!20] (b5) {Offer gen.\\automate};
\draw (rootB) -- (b1); \draw (rootB) -- (b2); \draw (rootB) -- (b3); \draw (rootB) -- (b4); \draw (rootB) -- (b5);
\node[below=0.15cm of b3, font=\scriptsize, align=center] {\textbf{Scenario B}: $\rho(P)=4/5=0.80$\\HITL anchor at underwriting $\Rightarrow$ \textbf{no warning}};
\end{tikzpicture}
\caption{Automation-debt warning, two scenarios on the same five-leaf loan-origination process. Left (A): all five subtasks classified automate; density saturates ($\rho=1.00$) and no leaf carries a regulator-mandated HITL anchor, so the warning fires. Right (B): underwriting is classified hybrid with an HITL anchor grounded in adverse-action explainability rules; the density drops to $\rho=0.80$ but the anchor neutralises the warning. Full mechanics in Appendix~\ref{app:debt}.}
\label{fig:debt}
\end{figure}
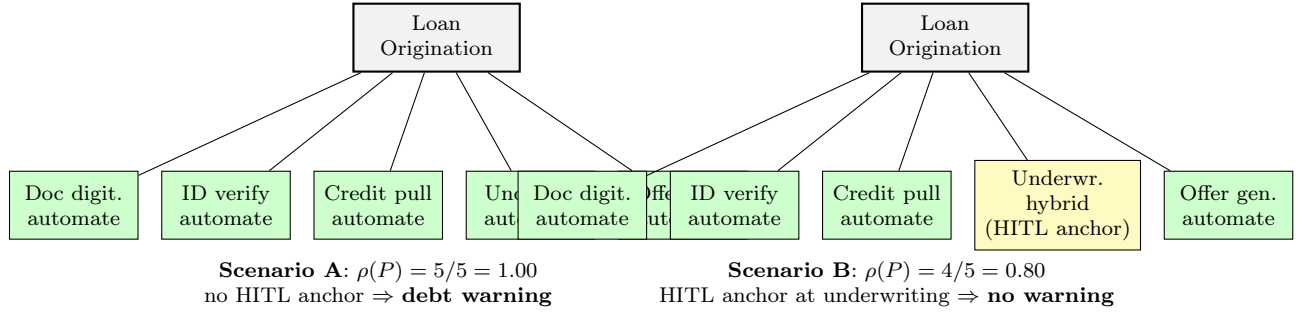

A process triggers an automation-debt warning when $\rho(P) \geq 0.80$ and no leaf subtask carries a regulator-mandated human-in-the-loop anchor ($\mathrm{hitl\_required\_by\_regulation} = \mathrm{true}$ for at least one augment, hybrid, or preserve leaf). The HITL-anchor exception encodes the principle that a process whose oversight is grounded in a specific regulatory requirement at any point retains a defensible recovery point even at high automation density. Figure~\ref{fig:debt} contrasts the two scenarios; full mechanics (hierarchical rollup, link to automation sovereignty) are in Appendix~\ref{app:debt}.

\begin{table}[htbp]
\centering
\caption{Four decision outcomes with prescribed governance actions.}
\label{tab:outcomes}
\begin{tabular}{@{}l p{4.3cm} p{6cm}@{}}
\toprule
\textbf{Decision} & \textbf{Meaning} & \textbf{Governance Action} \\
\midrule
automate & All gates cleared, composite $< 0.60$ & Proceed. Quarterly re-evaluation. \\
preserve & G1, G2, or G3 failed & Keep human role. Document rationale. Annual review. \\
augment & G5 failed & AI assists, human decides. Gradual knowledge capture. \\
hybrid & G4 or composite failed & AI executes, human supervises. Role splitting. \\
\bottomrule
\end{tabular}
\end{table}

\subsection{Four Decision Outcomes}

\paragraph{Time-indexed evaluation.} The risk profile of a decision is not static: regulation evolves, AI capabilities improve, knowledge can be partially documented over time, and reversibility costs grow as dependency deepens. The protocol projects each gate score $s_d(0) \in [0,1]$ over a decision horizon $H \in \{1,3,5\}$ years using a first-order linear approximation, clipped to the score domain:

\begin{equation}
s_d(H) = \mathrm{clip}\big(s_d(0) + \delta_d \cdot H,\ 0,\ 1\big)
\label{eq:drift}
\end{equation}

where $\delta_d$ is an annual drift rate per dimension. The financial-services default drifts are $\delta_{\mathrm{TKE}} = +0.03/\mathrm{yr}$ (knowledge atrophies as AI handles routine cases), $\delta_{\mathrm{RR}} = -0.02/\mathrm{yr}$ (AI reliability improves), $\delta_{\mathrm{RE}} = +0.05/\mathrm{yr}$ (regulatory frameworks converge toward mandatory oversight), and $\delta_{\mathrm{SCD}} = 0/\mathrm{yr}$ (trust effects stable in the medium term). The decision-flip horizon $H^* = \min\{H : \mathrm{decision}(s_d(H)) \neq \mathrm{decision}(s_d(0))\}$ is the smallest projected horizon at which the outcome inverts; $H^* = \infty$ if no horizon flips it. Illustrative example: a role currently passing G4 at $\mathrm{RE}(0) = 0.62$ projects under $\delta_{\mathrm{RE}} = +0.05$ to $\mathrm{RE}(3) = 0.77 > 0.70$, so $H^* = 3$ years and the role receives conditional automate---approved today with mandatory re-evaluation at $H^*$. The mirror case is deferred preserve: a role failing today on a dimension with $\delta_d < 0$ whose projected score crosses below threshold within 5 years is preserved now, with automation re-evaluation triggered when the projection crosses. Full drift-rate derivation and v0.2.0 operationalisation status are in Appendix~\ref{app:time-indexed}.

\section{Operationalization}
\label{sec:operationalization}

Operationalising the protocol imposes three constraints. First, scoring reduces to a 40-field structured input schema---this paper's v1 reference (Appendix~\ref{app:schema})---that maps every score sub-component of Section~\ref{sec:dimensions} to an observable, falsifiable field. The schema has seven functional groups: 6 task-context fields, 2 Gate 1 (net-benefit) fields, 8 TKE fields, 5 RR fields, 8 RE fields, 6 SCD fields, and 5 qualitative-context fields used for narrative justification but not for scoring. Second, scoring is deterministic over those inputs: large-language-model inference is restricted to assistive roles---decomposing a process description into discrete tasks, and producing post-scoring narrative justifications---and never enters the scoring path, preserving auditability under the EU AI Act traceability requirements (Art. 12 record-keeping for high-risk systems) \cite{euaiact2024} and NIST AI RMF traceability guidance (Measure function) \cite{nist2023airmf}. Third, determinism plus explicit configuration of thresholds and weights yields a fully reproducible audit trail traceable to (a) the structured inputs, (b) the gate scores and composite, (c) the configuration version (thresholds, weights, schema), and (d) the protocol version. Two evaluators given the same inputs obtain the same decision; a regulator inspecting a past decision can reconstruct it from artefacts alone. The full discussion of each constraint, including the rationale for excluding LLMs from the scoring path, is in Appendix~\ref{app:operationalization-detail}.

\paragraph{Schema authority and extensibility.} The 40-field schema reported here is the AI Lab Santander v1 schema for financial services. It is part of the configuration layer, not the protocol layer: institutions deploying PHP-AIO in other industries (healthcare, public administration, retail) are expected to extend or restrict the schema to fit their observable role attributes, subject to two invariants. Invariant 1 (coverage): every formula sub-component of Section~\ref{sec:dimensions} must remain populated by at least one field---extensions never weaken the score, only enrich it. Invariant 2 (versioning): schema changes are versioned and approver-signed in the same audit trail as threshold and weight changes (Appendix~\ref{app:operationalization-detail}), so reproducibility holds within a schema version. The financial-services v1 schema is offered as a reference; deviations are governance choices, not protocol violations.

\section{Practical Examples}
\label{sec:examples}

Table~\ref{tab:examples} reports four internal back-office processes yielding three distinct outcomes under PHP-AIO; Figure~\ref{fig:heatmap} renders the same evaluation as a per-cell heatmap that makes the deciding gate visually salient. None of the four involves external customers or regulator-mandated oversight; the protocol discriminates on tacit knowledge, resilience, and social-capital grounds alone. hybrid is a regulator-driven outcome (G4) and is therefore not represented in this set. The scores are illustrative of protocol behaviour on stylised role profiles; per-role detail (input rationale, gate-by-gate narrative) is in Appendix~\ref{app:per-role}.

\begin{table}[htbp]
\centering
\caption{Four roles, three distinct outcomes. `--' indicates the gate is not evaluated because the protocol stops at the first failure (sequential structure of Section~\ref{sec:protocol}). All four examples pass G1.}
\label{tab:examples}
\begin{tabular}{@{}l c c c c@{}}
\toprule
\textbf{Gate} & \textbf{Invoice Entry} & \textbf{Incident Triage} & \textbf{Mgr Coaching} & \textbf{Architecture Review} \\
\midrule
G1 Net Benefit & 50\% \checkmark & 50\% \checkmark & 50\% \checkmark & 50\% \checkmark \\
G2 TKE & 0.00 \checkmark & 0.08 \checkmark & 0.08 \checkmark & 0.80 $\times$ \\
G3 RR & 0.23 \checkmark & 0.82 $\times$ & 0.32 \checkmark & -- \\
G4 RE & 0.07 \checkmark & -- & 0.07 \checkmark & -- \\
G5 SCD & 0.13 \checkmark & -- & 0.83 $\times$ & -- \\
Composite & 0.10 \checkmark & -- & -- & -- \\
\midrule
\textbf{Outcome} & \textbf{automate} & \textbf{preserve} & \textbf{augment} & \textbf{preserve} \\
\bottomrule
\end{tabular}
\end{table}

PHP-AIO does not produce a single recommendation for all roles. The same organisation can automate invoice entry, preserve incident triage on resilience grounds, augment people-manager coaching on social-capital grounds, and preserve senior architecture reviews on tacit-knowledge grounds---four assessments yielding three distinct outcomes, each grounded in the gate that decided it.

\begin{table}[htbp]
\centering
\caption{Gate-by-gate evaluation of the four practical examples. Green = pass, red = failing gate (decides outcome), grey `--' = not evaluated (sequential stop).}
\label{fig:heatmap}
\begin{tabular}{@{}l c c c c c c l@{}}
\toprule
 & \textbf{G1} & \textbf{G2} & \textbf{G3} & \textbf{G4} & \textbf{G5} & \textbf{Composite} & \textbf{Outcome} \\
 & Net Ben. & TKE & RR & RE & SCD & & \\
\midrule
Invoice Entry & \cellcolor{green!25}50\% & \cellcolor{green!25}0.00 & \cellcolor{green!25}0.23 & \cellcolor{green!25}0.07 & \cellcolor{green!25}0.13 & \cellcolor{green!25}0.10 & automate \\
Incident Triage & \cellcolor{green!25}50\% & \cellcolor{green!25}0.08 & \cellcolor{red!25}0.82 & \cellcolor{gray!15}-- & \cellcolor{gray!15}-- & \cellcolor{gray!15}-- & preserve \\
Mgr Coaching & \cellcolor{green!25}50\% & \cellcolor{green!25}0.08 & \cellcolor{green!25}0.32 & \cellcolor{green!25}0.07 & \cellcolor{red!25}0.83 & \cellcolor{gray!15}-- & augment \\
Arch. Review & \cellcolor{green!25}50\% & \cellcolor{red!25}0.80 & \cellcolor{gray!15}-- & \cellcolor{gray!15}-- & \cellcolor{gray!15}-- & \cellcolor{gray!15}-- & preserve \\
\bottomrule
\end{tabular}
\end{table}

\section{Limitations}

The protocol presented here is a v1 conceptual framework. Five limitations should be made explicit.

\paragraph{Expert-elicited parameters.} All quantitative parameters---per-dimension weights, gate thresholds, country multipliers, and drift rates---were elicited by internal Santander AI Lab consensus drawing on the automation-bias and algorithmic-governance literature \cite{cummings2004automation, mittelstadt2019principles}, not optimised against ground-truth data. The sensitivity analysis in Appendix~\ref{app:calibration} shows single-gate decisions are robust to the implied uncertainty but composite-driven decisions are not. Refining multipliers against regulatory-fine datasets and drift rates against multi-year panels are tractable next steps.

\paragraph{Validation is ongoing.} The Section~\ref{sec:examples} examples are stylised role profiles; a retrospective validation study against historical automation decisions with observable outcomes (knowledge loss, regulatory action, trust failure) is currently underway, though no results are available yet. A multi-institution case panel spanning at least five years with matched outcome data remains the principal validation item and the precondition for moving PHP-AIO from candidate framework to evidenced framework.

\paragraph{Input-level inter-rater reliability is unmeasured.} The reproducibility guarantee of Section~\ref{sec:operationalization} is conditional: determinism holds from structured inputs to decision, not from role description to structured inputs. Several of the 40 schema fields are Likert judgments (\texttt{requires\_expert\_judgment}, \texttt{brand\_perception\_impact}, \texttt{team\_morale\_impact}) on which two independent analysts can plausibly diverge, and the protocol currently has no mechanism to detect such divergence. A two-rater study measuring agreement (Cohen's $\kappa$) on the input fields and on the resulting decisions over a shared role set is part of the validation work plan.

\paragraph{Single-industry calibration.} The reported configuration is financial-services-specific. The architecture is industry-agnostic but its quantitative defaults are not transferable without recalibration: healthcare, public administration, and retail each have qualitatively different RE structures and SCD weights. Reference configurations for those sectors are future work.

\paragraph{Dual-use risk and configuration audit.} Configurability is also the surface that lets an institution ratify cost-driven outcomes under procedural cover. Three mitigations apply: the financial-services configuration is the minimum-compliance baseline\footnote{Here minimum-compliance baseline denotes the internal baseline against which approved configuration deltas are audited at the operating institution; it is not a claim that these values constitute a regulatory floor or an industry standard.} against which reconfigurations are audited as approved deltas; every threshold and weight change is a versioned, timestamped, approver-signed artefact, so outcome-shopping cannot be disguised as parameter-tuning; and the retrospective validation mentioned above will yield empirical bounds within which protocol outcomes correlate with observed ones---deviations outside that band become out-of-distribution governance choices.

\section{Conclusion}

PHP-AIO formalises a four-dimension risk profile---tacit-knowledge erosion, resilience reduction, regulatory exposure, and socio-institutional capital degradation---as a deterministic five-gate protocol that yields an auditable automate / preserve / augment / hybrid decision before an organisation commits to deployment. It is not a competitor to NIST AI RMF, EU AI Act conformity assessment, FAccT-style impact assessments, or pre-deployment risk-profile disclosure standards \cite{sherman2024risk}---all of which act once a deployment candidate exists---but a complement upstream of them: a pre-decision filter whose output is a versioned configuration plus a gate-by-gate score, not a narrative.

The separation between deterministic scoring and optional LLM assistance is what makes the protocol auditable under EU AI Act traceability requirements (Art. 12 record-keeping) \cite{euaiact2024} and NIST AI RMF traceability guidance (Measure function) \cite{nist2023airmf}: identical inputs yield identical decisions regardless of model, prompt, or runtime, and every configuration change is a timestamped, approver-signed artefact. Per-domain configurability of weights, thresholds, and country multipliers lets each institution recalibrate to its own risk appetite, with the financial-services configuration in this paper offered as a v1 expert-elicited baseline rather than a universal default.

A retrospective validation against documented automation decisions is currently underway; together with cross-industry calibration for healthcare and public administration and learning drift rates from accumulated assessments, these are the three lines of work that would move PHP-AIO from a v1 governance protocol toward an institution-specific, evidenced risk model. The automation decision is too consequential to leave to cost savings alone; PHP-AIO is offered as one way to ensure it is not.

\bibliographystyle{plain}
\bibliography{references}

\appendix

\section{Input Schema}
\label{app:schema}

The reference implementation operationalises the protocol with a 40-field input schema. Six fields capture task context (stored at the task level), 29 fields drive the five gates (stored at the assessment level), and 5 fields provide qualitative context used by the narrative layer but not by scoring. Field names match the implementation's database schema. This is the financial-services v1 schema; extensions for other industries follow the schema-authority invariants stated in Section~\ref{sec:operationalization}.

\subsection{Task Context (6 fields)}
\begin{tabular}{@{}l l l@{}}
\toprule
\textbf{Field} & \textbf{Type} & \textbf{Values / Range} \\
\midrule
name & string & free text \\
description & text & free text \\
task\_type & enum & decision, validation, data\_entry, \ldots \\
frequency & enum & per\_case, daily, weekly, monthly, ad-hoc \\
avg\_time\_minutes & int & $\geq 0$ \\
criticality & enum & blocking, important, optional \\
\bottomrule
\end{tabular}

\subsection{Gate 1 --- Net Benefit (2 fields)}
Gate 1 derives $\mathrm{efficiency\_gain} = (\mathrm{current} - \mathrm{ai})/\mathrm{current}$ and tests it against the $\geq 15\%$ threshold.

\begin{tabular}{@{}l l l@{}}
\toprule
\textbf{Field} & \textbf{Type} & \textbf{Range} \\
\midrule
current\_cost\_per\_case & float & $\geq 0$ \\
estimated\_ai\_cost & float & $\geq 0$ \\
\bottomrule
\end{tabular}

\subsection{Gate 2 --- TKE (8 fields)}
\begin{tabular}{@{}l l l l@{}}
\toprule
\textbf{Field} & \textbf{Type} & \textbf{Range / Values} & \textbf{Drives} \\
\midrule
requires\_expert\_judgment & int & 1--5 (Likert) & $\kappa$ \\
exception\_frequency & float & 0--100\% & $\kappa$ \\
documentation\_level & int & 1--5 (Likert) & $\kappa$ \\
min\_experience\_years & int & $\geq 0$ & IR \\
training\_time\_months & int & $\geq 0$ & IR \\
knowledge\_concentration & bool & true / false & IR \\
downstream\_dependencies & list & task ids & CM \\
error\_impact & enum & low / medium / high / critical & CM \\
\bottomrule
\end{tabular}

\subsection{Gate 3 --- RR (5 fields)}
\begin{tabular}{@{}l l l l@{}}
\toprule
\textbf{Field} & \textbf{Type} & \textbf{Range / Values} & \textbf{Drives} \\
\midrule
manual\_backup\_exists & bool & true / false & RC \\
max\_tolerable\_downtime\_hours & int & $\geq 0$ & RC \\
detects\_novel\_threats & bool & true / false & AD \\
escalation\_rate & float & 0--100\% & AD \\
dependent\_automated\_systems & list & system identifiers & CR \\
\bottomrule
\end{tabular}

\subsection{Gate 4 --- RE (8 fields)}
\begin{tabular}{@{}l l l l@{}}
\toprule
\textbf{Field} & \textbf{Type} & \textbf{Range / Values} & \textbf{Drives} \\
\midrule
affects\_customers\_directly & bool & true / false & CQ \\
involves\_credit\_decisions & bool & true / false & CQ \\
handles\_sensitive\_data & bool & true / false & CQ \\
subject\_to\_audit & list & audit body codes & JU \\
hitl\_required\_by\_regulation & bool & true / false & JU \\
regulation\_reference & string & free text (e.g.\ EU AI Act Art.\ 14) & JU \\
precedent\_sanctions & bool & true / false & JU \\
estimated\_reversal\_cost & float & EUR, $\geq 0$ & RV \\
\bottomrule
\end{tabular}

\subsection{Gate 5 --- SCD (6 fields)}
\begin{tabular}{@{}l l l l@{}}
\toprule
\textbf{Field} & \textbf{Type} & \textbf{Range / Values} & \textbf{Drives} \\
\midrule
customer\_interaction\_type & enum & none, email, phone, video, in\_person & RL \\
customer\_expects\_human & bool & true / false & RL \\
brand\_perception\_impact & int & 1--5 (Likert) & LG \\
publicly\_visible\_role & bool & true / false & LG \\
team\_morale\_impact & int & 1--5 (Likert) & LB \\
employees\_in\_role & int & $\geq 0$ & LB \\
\bottomrule
\end{tabular}

\subsection{Qualitative Context (5 fields, non-scoring)}
These fields enrich the LLM-generated narrative justification and the audit trail; they do not enter any gate formula.

\begin{tabular}{@{}l l@{}}
\toprule
\textbf{Field} & \textbf{Type} \\
\midrule
additional\_context & free text \\
known\_risks & free text \\
previous\_automation\_attempts & free text \\
stakeholder\_concerns & free text \\
notes & free text \\
\bottomrule
\end{tabular}

\section{Detailed Worked Examples}
\label{app:worked-examples}

This appendix expands the one-line worked examples of Section~\ref{sec:dimensions} into full derivations and ties each dimension to the broader literature on sovereignty degradation.

\subsection{Tacit Knowledge Erosion (TKE)}

A role requiring 5 years of experience, high expert judgment, and high error impact yields normalised sub-scores $\kappa = 0.41$, $\mathrm{IR} = 0.44$, $\mathrm{CM} = 0.64$ (raw values 0.49, 0.50, 0.69 rescaled from their effective ranges $[0.14, 1]$, $[0.10, 1]$, $[0.15, 1]$), hence $\mathrm{TKE} = 0.4 \cdot 0.59 + 0.3 \cdot 0.44 + 0.3 \cdot 0.64 \approx 0.56$---below the 0.70 Gate 2 threshold but flagged at the composite gate. TKE measures sovereignty degradation through the irreversible loss of judgment that cannot be re-encoded once the role is removed.

\subsection{Resilience Reduction (RR)}

A loan-approval role with no manual fallback and 24-hour tolerance (RC raw 0.60, normalised 0.64), no novel-threat detection and 20\% escalation (AD raw 0.56, normalised 0.58), and three dependent downstream systems ($\mathrm{CR} = 0.60$, already in $[0,1]$) yields $\mathrm{RR} = 0.64 \times 0.35 + 0.58 \times 0.35 + 0.60 \times 0.30 \approx 0.61$---below the 0.70 threshold, flagged at the composite gate.

This dimension draws on antifragility theory \cite{taleb2012antifragile} and on the broader literature on automation-induced complacency in safety-critical contexts \cite{cummings2004automation}: diverse response mechanisms outperform purely automated systems under novel or adversarial conditions. RR measures sovereignty degradation through the structural fragility introduced when correlated automated systems share failure modes.

\subsection{Regulatory Exposure (RE)}

A UK loan-approval role (affects customers directly, credit decision, sensitive data, three audit bodies, HITL mandatory, precedent sanctions, \texteuro200K reversal cost) yields raw values $\mathrm{CQ} = 0.84$, $\mathrm{JU} = 1.034$, $\mathrm{RV} = 0.40$ which rescale to 1.000, 0.913 and 0.400 respectively, hence $\mathrm{RE} = 1.000 \times 0.4 + 0.913 \times 0.4 + 0.400 \times 0.2 = 0.4 + 0.3652 + 0.08 = 0.8452 \approx 0.85$---above the 0.70 threshold, so the protocol routes the role to hybrid.

For multi-jurisdictional organisations operating across heterogeneous regulatory regimes, this dimension captures the country-level variation that single-jurisdiction analyses miss. RE measures sovereignty degradation through the regulatory option value lost when an automated decision is later mandated to be reversed.

\subsection{Socio-Institutional Capital Degradation (SCD)}

A retail in-branch advisor role (in-person interaction, customers expect a human, brand impact 5/5, publicly visible, 30 employees, morale impact 4/5) yields raw values $\mathrm{RL} = 0.92$, $\mathrm{LG} = 0.92$, $\mathrm{LB} = 0.70$ which rescale to 1.00, 1.00 and 0.67 respectively, hence $\mathrm{SCD} = 1.00 \times 0.35 + 1.00 \times 0.35 + 0.67 \times 0.30 \approx 0.90$---above the 0.70 threshold, so the protocol routes the role to augment (AI assists the advisor, the human keeps the relationship). SCD measures sovereignty degradation through the trust capital depleted when consequential interactions lose their human anchor.

\section{Sub-Dimension Construction}
\label{app:subdimensions}

This appendix gives the raw-input construction for each of the twelve sub-dimensions in Section~\ref{sec:dimensions}. All sub-dimensions are subsequently affinely rescaled to $[0,1]$ following the common-scale convention.

\subsection{TKE sub-dimensions}
\begin{itemize}[leftmargin=*]
\item \textbf{Codifiability} ($\kappa$) $\in [0.14, 1]$: what fraction of the role's judgment can be formally represented. 40\% from expert-judgment requirement on a 1--5 Likert scale (normalised as $(6 - ej)/5$, so $ej=1 \rightarrow 1.0$, $ej=5 \rightarrow 0.2$), 30\% from the complement of exception frequency ($1 - exc/100$), 30\% from documentation level on a 1--5 scale ($doc/5$). A fully documented role with no exceptions and no expert judgment saturates at $\kappa=1$; a role with $ej=5$, $exc=100\%$, $doc=1$ hits the raw minimum $0.4 \cdot 0.2 + 0.3 \cdot 0 + 0.3 \cdot 0.2 = 0.14$. The TKE formula uses $(1-\kappa)$ so that lower codifiability raises risk.
\item \textbf{Irreversibility} (IR) $\in [0.10, 1]$: 40\% from minimum experience required ($\min(years, 10)/10$), 40\% from training time ($\min(months, 24)/24$), 20\% from knowledge concentration (1.0 if concentrated, 0.5 if not). Years are capped at 10 and training at 24 months. A role with 5 years' experience, 6 months' training, knowledge concentrated in a few people scores raw $0.4 \cdot 0.5 + 0.4 \cdot 0.25 + 0.2 \cdot 1.0 = 0.50$; a junior role with no training and distributed knowledge hits raw $0.2 \cdot 0.5 = 0.10$.
\item \textbf{Criticality} (CM) $\in [0.15, 1]$: 60\% from error-impact severity (low $\rightarrow 0.25$, medium $\rightarrow 0.50$, high $\rightarrow 0.75$, critical $\rightarrow 1.0$), 40\% from downstream dependencies $\min(n/5, 1)$. A high-impact role with three dependent tasks scores raw $0.6 \cdot 0.75 + 0.4 \cdot 0.6 = 0.69$; a low-impact role with no dependencies hits raw $0.6 \cdot 0.25 = 0.15$.
\end{itemize}

\subsection{RR sub-dimensions}
\begin{itemize}[leftmargin=*]
\item \textbf{Recovery} (RC) $\in [0,1]$: 50\% from whether a manual backup process exists (0.3 if yes, 0.7 if not) and 50\% from downtime tolerance ($1 - \min(hours/48, 1)$). A role with no manual backup and 1-hour tolerance scores $\approx 0.84$; a role with backup and 48-hour tolerance scores 0.15.
\item \textbf{Adversarial} (AD) $\in [0,1]$: 60\% from whether the role detects novel threats today (0.2 if yes, 0.8 if not) and 40\% from the human-escalation rate (0--100\%, normalised to 0--1). A role that catches no novel threats and escalates 30\% of cases scores $0.8 \times 0.6 + 0.30 \times 0.4 = 0.60$.
\item \textbf{Correlation} (CR) $\in [0,1]$: number of dependent automated systems divided by 5, capped at 1. Five or more correlated systems saturate this term at 1.0.
\end{itemize}

\subsection{RE sub-dimensions}
\begin{itemize}[leftmargin=*]
\item \textbf{Consequentiality} (CQ) $\in [0,1]$: 40\% from whether the decision affects customers directly (0.8 if yes, 0.2 if not), 40\% from whether it involves credit decisions (1.0 if yes, 0.0 if not), 20\% from whether it handles sensitive personal data (0.6 if yes, 0.2 if not). A retail credit decision touching sensitive data reaches the raw maximum 0.84, which rescales to 1.0.
\item \textbf{Jurisdictional} (JU) $\in [0,1]$: an inner score (30\% from audit bodies $\min(n/3, 1)$, 40\% from whether HITL is required by regulation (1.0 if yes, 0.0 if not), 30\% from precedent sanctions (0.8 if yes, 0.2 if not)) multiplied by a country factor: 1.2$\times$ for EU AI Act jurisdictions (Spain, Germany, Portugal, Poland), 1.1$\times$ for the UK (FCA/PRA), 1.0$\times$ for the US, 0.9$\times$ for Brazil and Mexico, 0.8$\times$ for Argentina and Chile. The asymmetric mapping for HITL (1.0/0.0, not 0.8/0.2) reflects the fact that a legally-binding HITL requirement is a hard regulatory constraint, not a soft signal. The raw product spans $[0.048, 1.128]$ and is then rescaled to $[0,1]$, so the country multiplier preserves its directional effect even at the upper saturation. A UK role with three audit bodies, HITL required, and precedent sanctions has raw $(0.3 \cdot 1 + 0.4 \cdot 1 + 0.3 \cdot 0.8) \times 1.1 = 0.94 \times 1.1 = 1.034$, which rescales to $\approx 0.91$.
\item \textbf{Reversal} (RV) $\in [0,1]$: estimated cost to reintroduce human oversight, divided by \texteuro500K, capped at 1. A \texteuro200K reversal cost scores 0.4; \texteuro500K or more saturates at 1.0.
\end{itemize}

\subsection{SCD sub-dimensions}
\begin{itemize}[leftmargin=*]
\item \textbf{Relational} (RL) $\in [0,1]$: 60\% from interaction type (none 0.0, email 0.2, phone 0.5, video 0.7, in-person 1.0) and 40\% from whether the customer expects a human (0.8 if yes, 0.2 if not). An in-person role with explicit customer expectation reaches the raw maximum 0.92 (rescaled to 1.0); a fully back-office role hits raw 0.08 (rescaled to 0).
\item \textbf{Legitimacy} (LG) $\in [0,1]$: 60\% from brand-perception impact (1--5 scale, normalised by 5) and 40\% from whether the role is publicly visible (0.8 if yes, 0.2 if not). A branch manager (impact 5, publicly visible) reaches the raw maximum 0.92 (rescaled to 1.0); a back-office data-entry role (impact 1, not visible) hits raw 0.20 (rescaled to 0).
\item \textbf{Labor} (LB) $\in [0,1]$: 50\% from team-morale impact (1--5 scale, normalised by 5) and 50\% from headcount in the role ($\min(n/50, 1)$). A role held by 30 people with morale impact 4/5 has raw $0.8 \times 0.5 + 0.6 \times 0.5 = 0.70$ (rescaled to $\approx 0.67$); 50+ people with maximum morale impact saturates at 1.0.
\end{itemize}

\section{Threshold Calibration and Sensitivity}
\label{app:calibration}

The gate thresholds in Table~\ref{tab:gates} are derived from three converging sources: regulatory risk-tolerance standards in financial services, the asymmetric cost structure of irreversible decisions, and robustness analysis of the four practical examples in Section~\ref{sec:examples}.

\paragraph{Regulatory grounding.} The EU AI Act \cite{euaiact2024} classifies credit scoring, insurance risk assessment, and employment decisions as high-risk AI applications requiring mandatory human oversight. EBA Guidelines on internal governance \cite{eba2021guidelines} require documented governance and accountability arrangements over institutions' risk-taking decisions, including those that are supported or executed by automation. The RE threshold of 0.70 is calibrated to this ceiling: scores above 0.70 correspond to roles where automated operation would conflict with existing or imminent regulatory requirements in at least one of the five jurisdictions where the protocol is currently being piloted (a subset of the ten countries with explicit jurisdictional multipliers in the RE formula).

\paragraph{Irreversibility asymmetry.} Each of the four risk gates protects against an outcome whose recovery cost is asymmetric: tacit knowledge once eliminated cannot be reconstructed on demand (G2), organisational resilience once reduced cannot be restored instantly during a failure event (G3), regulatory non-compliance once incurred cannot be undone retroactively (G4), and trust capital once lost takes years to rebuild (G5). In all four cases preservation is reversible while the failure mode is not, which justifies conservative thresholds. The uniform value 0.70 is set at the point above which the underlying sub-dimension combinations produce damage exceeding what structured documentation, redundancy, regulatory remediation, or trust-restoration programmes have historically been able to recover \cite{collins2010tacit, nonaka1995knowledge}.

\paragraph{Sensitivity analysis.} Table~\ref{tab:sensitivity} reports the binding margin for each outcome in Section~\ref{sec:examples}. All four examples tolerate upward threshold perturbations of at least $+14\%$ before the outcome inverts; downward perturbations are unbounded for the failing-gate cases, since they only reinforce the decision. The narrowest binding margin is $+0.10$ absolute (Architecture Review, TKE). Decisions driven by a single failed gate are therefore robust under the threshold uncertainty inherent to a v1 framework. Composite-driven outcomes are weight-sensitive by construction; the explicit margin to the composite threshold of 0.60 is reflected in the table's first row. For Invoice Data Entry the margin admits a stronger statement: the composite is a convex combination of the four dimension scores $\{0.00, 0.23, 0.07, 0.13\}$, so no weight vector summing to one can lift it above the maximum dimension score of 0.23, far below the 0.60 threshold. The automate outcome is therefore invariant to any reweighting as well as to threshold shifts; composite-driven outcomes closer to the threshold do not enjoy this bound.

\paragraph{Configurability.} The defaults reported in Table~\ref{tab:gates} are the financial-services configuration. Every threshold is configurable per domain, jurisdiction, and risk appetite; thresholds are part of the configuration that an institution must own and document.

\begin{table}[htbp]
\centering
\caption{Threshold sensitivity for the four examples of Section~\ref{sec:examples}. Tolerable upward shift is the maximum proportional increase in the deciding-gate threshold that preserves the outcome.}
\label{tab:sensitivity}
\begin{tabular}{@{}l l c c c l@{}}
\toprule
\textbf{Role} & \textbf{Deciding gate} & \textbf{Score} & \textbf{Threshold} & \textbf{Margin} & \textbf{Tolerable upward shift} \\
\midrule
Invoice Data Entry & Composite & 0.10 & 0.60 & $-0.50$ & unbounded \\
Incident Triage & RR & 0.82 & 0.70 & $+0.12$ & $+17.1\%$ \\
Manager Coaching & SCD & 0.83 & 0.70 & $+0.13$ & $+18.6\%$ \\
Architecture Review & TKE & 0.80 & 0.70 & $+0.10$ & $+14.3\%$ \\
\bottomrule
\end{tabular}
\end{table}

\section{Automation Debt: Worked Example and Operationalisation}
\label{app:debt}

\paragraph{Worked example.} Consider a loan-origination process with five leaf subtasks---document digitisation, identity verification, credit-bureau pull, underwriting model scoring, and offer generation---with approximately equal average handling times so that $w_i$ is uniform and $\rho(P)$ reduces to the share of automate leaves.

\paragraph{Scenario A (warning triggered).} All five subtasks score automate individually. Then $\rho(P) = 5/5 = 1.00 \geq 0.80$, and no leaf carries a regulator-mandated HITL anchor. The process triggers an automation-debt warning that individual gate analysis would have missed: five locally safe decisions compose into a fully automated end-to-end chain with no recovery point.

\paragraph{Scenario B (saved by the HITL anchor).} Same five subtasks, but underwriting model scoring is classified hybrid with $\mathrm{hitl\_required\_by\_regulation} = \mathrm{true}$ (e.g.\ adverse-action explainability under consumer-credit rules). Then $\rho(P) = 4/5 = 0.80$, and the HITL-anchor condition is satisfied at the underwriting leaf. No warning fires: the regulator-grounded oversight point is treated as a structural recovery mechanism that defeats the density signal. This is the mechanic by which regulatory grounding does real work in the protocol.

\paragraph{Hierarchical rollup mechanics.} $\rho$ is computed at every parent task in the process tree, not only at the root: each non-leaf node aggregates over the leaves of its own subtree using the same time-weighted formula, so warnings can surface at any level of the decomposition. A departmental sub-process whose leaves are all automated triggers locally even if the enclosing process as a whole stays under threshold.

\paragraph{Relationship to automation sovereignty.} Automation debt is the process-level measure of sovereignty degradation: $\rho(P)$ quantifies how much of a process an organisation can no longer operate manually, and the HITL-anchor exception encodes the requirement that high density is tolerable only when at least one regulator-grounded human intervention point remains. The signal is intentionally density-only at this stage; richer risk-weighted formulations are a calibration question for the empirical-validation work plan.

\section{Time-Indexed Protocol}
\label{app:time-indexed}

The protocol as defined in Sections~\ref{sec:dimensions}--\ref{sec:protocol} evaluates a role at a point in time. The risk profile of an automation decision is not static: regulatory requirements evolve (the EU AI Act phases into full application through 2026), AI capabilities improve, tacit knowledge can be partially documented over time, and the reversibility cost grows as the organisation's dependency on the automated system deepens.

PHP-AIO supports time-indexed evaluation by projecting each gate score over a decision horizon $H \in \{1,3,5\}$ years using a first-order linear approximation, clipped to the score domain:

\begin{equation}
s_d(H) = \mathrm{clip}\big(s_d(0) + \delta_d \cdot H,\ 0,\ 1\big)
\label{eq:drift-appendix}
\end{equation}

where $s_d(0) \in [0,1]$ is the baseline score on dimension $d \in \{\mathrm{TKE}, \mathrm{RR}, \mathrm{RE}, \mathrm{SCD}\}$ and $\delta_d$ is an annual drift rate (positive: risk increasing; negative: risk decreasing). The clip operator guarantees $s_d(H) \in [0,1]$ for any horizon and drift; non-linear refinements (logistic projection, saturating piecewise forms) are deferred to future work.

\paragraph{Default drift rates.} Table~\ref{tab:drift} reports indicative annual drifts for the financial-services configuration in the 2025--2027 regulatory phase-in window. These values are configurable per dimension and jurisdiction; empirical calibration of drift rates against multi-year observational data is part of the validation work plan.

\begin{table}[htbp]
\centering
\caption{Default annual drift rates for the financial-services configuration.}
\label{tab:drift}
\begin{tabular}{@{}l l p{7cm}@{}}
\toprule
\textbf{Dimension} & \textbf{Default $\delta$} & \textbf{Driver} \\
\midrule
TKE & $+0.03$/yr & Knowledge atrophies as AI handles routine cases; exception judgment becomes rarer and harder to rebuild. \\
RR & $-0.02$/yr & AI system reliability improves; manual fallback procedures, when actively maintained, mitigate dependency risk. \\
RE & $+0.05$/yr & Regulatory frameworks converge toward mandatory oversight; jurisdictional exposure increases. \\
SCD & $0.00$/yr & Trust effects stable in the medium term; institutional reputation changes slowly. \\
\bottomrule
\end{tabular}
\end{table}

\paragraph{Decision-flip horizon.} The most actionable derived metric is the decision-flip horizon $H^* = \min\{H \in \{1,3,5\} : \mathrm{decision}(s_d(H)) \neq \mathrm{decision}(s_d(0))\}$: the smallest projected horizon at which the gate outcome inverts (defining $H^* = \infty$ when no horizon flips the decision). A role currently classified automate with $H^* = 3$ years is decision-fragile under expected drift, even though it passes today.

\paragraph{Two derived decision types.} The time-indexed protocol extends the four base decisions of Table~\ref{tab:outcomes} with two horizon-aware variants:
\begin{itemize}[leftmargin=*]
\item \textbf{Conditional automate}: gate-clear today with $H^* < \infty$. Automation is approved today with a mandatory re-evaluation scheduled at $H^*$.
\item \textbf{Deferred preserve}: gate-failed today on a dimension whose $\delta_d < 0$, with $s_d(H) < \tau_d$ projected for some $H \leq 5$. Preservation is maintained now, automation re-evaluation triggers when the projected score falls below threshold.
\end{itemize}

Both extended outcomes are auditable: the decision record includes the projected score trajectory $\{s_d(0), s_d(1), s_d(3), s_d(5)\}$ and the re-evaluation trigger date.

\paragraph{Operationalisation status.} The time-indexed protocol is formalised here as part of the v1 protocol specification but is not yet operationalised in the reference implementation (v0.2.0). The implementation evaluates each role at a single point in time; horizon projection, the decision-flip horizon $H^*$, and the two derived decision types are deferred to a future iteration. Empirical calibration of the drift rates against multi-year panel data, and the integration of the projection into the assessment pipeline, are jointly part of the validation work plan.

\section{Operationalization Detail}
\label{app:operationalization-detail}

A protocol that exists only on paper does not change automation decisions. Translating the four-dimension formalism of Section~\ref{sec:dimensions} and the gate logic of Section~\ref{sec:protocol} into a procedure that a non-research user can execute consistently imposes three design constraints. These constraints are illustrated by a reference implementation deployed for empirical validation; the implementation is one of several possible instantiations and is not part of the protocol's specification.

\paragraph{Input schema and observability.} Each gate must reduce to a finite set of observable, falsifiable inputs. PHP-AIO formalises this requirement through a 40-field input schema (Appendix~\ref{app:schema}) that maps every score sub-component of Section~\ref{sec:dimensions} to a structured field elicitable from process documentation, role specifications, and stakeholder interviews. The schema has seven functional groups: 6 task-context fields; 2 Gate 1 cost fields; 8 TKE fields; 5 RR fields; 8 RE fields; 6 SCD fields; and 5 qualitative-context fields used for narrative justification but not for scoring. The schema is the contract between the protocol's mathematics and operational reality: no score is computable without all relevant fields, and every scoring field maps to a specific sub-component of one of the four risk dimensions.

\paragraph{Separation of concerns: deterministic scoring with stochastic assistance.} A central design choice is to keep large-language-model (LLM) inference entirely outside the scoring path. The LLM, where used, has two assistive roles: (i) decomposing a process description (e.g., Corporate Client Onboarding) into discrete tasks for independent assessment, and (ii) producing the narrative justification attached to each decision after scoring. Gate scores and the composite are computed deterministically from the structured inputs by the formulas of Section~\ref{sec:dimensions}; identical inputs yield identical decisions, regardless of model, prompt, or runtime. This separation is deliberate: it keeps the protocol auditable under the EU AI Act traceability requirements (Art. 12 record-keeping for high-risk systems) \cite{euaiact2024} and NIST AI RMF traceability guidance (Measure function) \cite{nist2023airmf}, while still permitting LLM assistance where stochasticity is acceptable---decomposition is editable by the user before scoring, narratives are explanatory rather than decisive.

\paragraph{Auditability and reproducibility.} Determinism over the input schema, combined with explicit configuration of thresholds and weights, yields a fully reproducible audit trail. Every decision is traceable to (a) the structured inputs that produced it, (b) the gate scores and composite computed from those inputs, (c) the configuration version (thresholds, weights, schema) active at the time, and (d) the protocol version. Two evaluators given the same inputs obtain the same decision; the same evaluator on different days obtains the same decision; a regulator inspecting a past decision can reconstruct it from artefacts alone. This property is the operational counterpart of the deterministic-scoring constraint above and the ultimate justification for it: any framework operating under regulated AI oversight regimes must produce decisions that survive ex-post inspection \cite{mitchell2019model}.

\section{Per-Role Worked Examples}
\label{app:per-role}

This appendix expands each row of Table~\ref{tab:examples} into the gate-by-gate evaluation and the input rationale that produces the score.

\subsection{Invoice Data Entry $\rightarrow$ automate}
\begin{tabular}{@{}l c c l@{}}
\toprule
\textbf{Gate} & \textbf{Score} & \textbf{Threshold} & \textbf{Result} \\
\midrule
G1 Net Benefit & 50\% & $\geq 15\%$ & PASS \\
G2 TKE & 0.00 & $< 0.70$ & PASS \\
G3 RR & 0.23 & $< 0.70$ & PASS \\
G4 RE & 0.07 & $< 0.70$ & PASS \\
G5 SCD & 0.13 & $< 0.70$ & PASS \\
Composite & 0.10 & $< 0.60$ & PASS \\
\bottomrule
\end{tabular}

Highly codifiable, fully documented, no expert judgment, manual fallback exists, no relational capital. Safe to automate end-to-end.

\subsection{Production Incident Triage $\rightarrow$ preserve}
\begin{tabular}{@{}l c c l@{}}
\toprule
\textbf{Gate} & \textbf{Score} & \textbf{Threshold} & \textbf{Result} \\
\midrule
G1 Net Benefit & 50\% & $\geq 15\%$ & PASS \\
G2 TKE & 0.08 & $< 0.70$ & PASS \\
G3 RR & 0.82 & $< 0.70$ & FAIL \\
\bottomrule
\end{tabular}

The triage role itself is moderately codifiable (G2 passes), but it is the organisation's last line of defence when production breaks: no manual fallback, one-hour tolerance for downtime, five downstream automated systems depend on it. Automating it creates correlated failure modes precisely when the rest of the stack is also failing.

\subsection{People-Manager Coaching $\rightarrow$ augment}
\begin{tabular}{@{}l c c l@{}}
\toprule
\textbf{Gate} & \textbf{Score} & \textbf{Threshold} & \textbf{Result} \\
\midrule
G1 Net Benefit & 50\% & $\geq 15\%$ & PASS \\
G2 TKE & 0.08 & $< 0.70$ & PASS \\
G3 RR & 0.32 & $< 0.70$ & PASS \\
G4 RE & 0.07 & $< 0.70$ & PASS \\
G5 SCD & 0.83 & $< 0.70$ & FAIL \\
\bottomrule
\end{tabular}

Internal one-on-one coaching where managers expect a human counterpart, the role anchors the firm's people-development brand, and 50 employees occupy similar positions. AI augments the coach (preparation, summaries, follow-up tracking) but the human owns the relationship.

\subsection{Senior Architecture Reviews $\rightarrow$ preserve}
\begin{tabular}{@{}l c c l@{}}
\toprule
\textbf{Gate} & \textbf{Score} & \textbf{Threshold} & \textbf{Result} \\
\midrule
G1 Net Benefit & 50\% & $\geq 15\%$ & PASS \\
G2 TKE & 0.80 & $< 0.70$ & FAIL \\
\bottomrule
\end{tabular}

Senior architects approving major design decisions: 80\% of cases are exceptions, documentation lags reality, ten years of context required, knowledge concentrated in a handful of people, critical error impact. The judgment cannot be codified without losing the judgment itself.

\section*{Declaration of Generative AI and AI-Assisted Technologies in the Writing Process}

During the preparation of this work the authors used Cascade (Windsurf's AI coding assistant, powered by Claude, Anthropic) for editorial review of structure and prose, consistency checking of mathematical formulas and worked numerical examples, alignment between the paper text and the reference implementation, and bibliography curation. All AI-generated suggestions were reviewed and edited by the authors. The authors take full responsibility for the content of the publication, including the formal definitions, gate-threshold calibrations, and the empirical claims that follow once the validation work plan is executed.

\end{document}